\documentclass[12pt, letterpaper]{article}
\usepackage{amsmath,amssymb,amsfonts}
\usepackage{algorithmic}
\usepackage{algorithm}
\usepackage{array}
\usepackage[caption=false,font=normalsize,labelfont=sf,textfont=sf]{subfig}
\usepackage{textcomp}
\usepackage{stfloats}
\usepackage{url}
\usepackage{verbatim}
\usepackage{float}
\usepackage{graphicx}
\usepackage{multirow}
\usepackage{dirtytalk}
\usepackage{xcolor}
\usepackage{cite}
\begin{document}
\title{A New 22 nm ULPLS Architecture to Detect 70 mV Minimum Input, Suitable for IOT Applications\vspace{-0.5cm}}
%%%%%%%%%%%%%%%%%%%%%%%%%%%%%%%%%%% affiliations %%%%%%%%%%%%%%%%%%%%%%%%%%%%%%%%%%%%%%%%%
\author{Jehan~Taraporewalla\textsuperscript{$\dagger$} and Dipankar~Saha\textsuperscript{$\dagger$,$\ddagger$,*}}
        % <-this % stops a space
\date{\textsuperscript{$\dagger$}Department of Electronics and Telecommunication Engineering,\\Indian Institute of Engineering Science and Technology Shibpur, Howrah-711103, India.\\
\textsuperscript{$\ddagger$}Department of Electronics and Communication Engineering,\\ Institute of Engineering and Management, Saltlake, Kolkata-700091, India.\\
\textsuperscript{*}University of Engineering and Management (UEM), Kolkata, India.\\
(Email: pc.jehan@gmail.com and dipsah\_etc@yahoo.co.in) \vspace{-0.7cm}}

%%%%%%%%%%%%%%%%%%%%%%%%%%%%%%%%%%%%%%%%%%%%%%%%%%%%%%%%%%%%%%%%%%%%%%%%%%%%
\maketitle
\begin{abstract}
Modern applications such as energy harvesting, signal monitoring in bio-medical sensing, portable point of care devices, etc. which involve state of the art mixed signal subsystems require robust ultra low power operation. Here in this work, a novel ultra low power level shifter (ULPLS) is proposed for sensing voltage signals in sub-threshold region. The proposed architecture is implemented in 22 nm technology using a dual power supply. The high and low supply voltages (V\textsubscript{ddH} \& V\textsubscript{ddL}) are set as 0.8 V and 0.4 V respectively. The key design features of ULPLS include a current limiting PMOS diode, a voltage divider, and an enhanced pull up network. The ULPLS exhibits a low power dissipation of $\sim$ 22.84 nW with a minimum $\sim$ 70 mV detection of input signal. The robustness of the design has been examined via worst case and Monte Carlo analyses. 
\end{abstract}


%%%%%%%%%%%%%%%%%%%%%%%%%%%%%%%%%%%%%%%%%%%%%%%%%%%%%%%%%%%%%%%%%%%%%%%%%%%%%%%%%%%%%%%%%%%%%%%%%%%%%%%%%%%%%%%%%%%%%%%%%%

\section{Introduction}
The major challenge in realizing digital circuits for various applications viz. signal monitoring \cite{IOT}, application specific system processor \cite{cascaded_LS,Osaki,Rajendran}, low voltage detection \cite{Matsuzka,Matsuzka_2}, and energy harvesting \cite{Energy_Harvesting} is ensuring ultra low power operation with improved noise immunity and robustness \cite{Low_power}. In order to enhance the energy efficiency of the circuits used for afore mentioned applications, we may opt for sub-threshold or near threshold design techniques \cite{CMOS_Scaling,Scaling_Wang,Saha}.\\

Voltage scaling is a routinely adopted solution for reducing power dissipation in modern mixed signal sub-systems \cite{Ajit}. Besides, the use of dual supply voltages (V\textsubscript{ddH} \& V\textsubscript{ddL}) can be useful for tuning the performance with superior energy efficiency \cite{Saha,Scaling_Wang}. The elementary component for such sub-threshold or near threshold dual V\textsubscript{dd} digital design is the level shifter (LS)  which ensures an effective communication between the low and high supply voltage domains \cite{Ajit,Usami}. The most widely used conventional LS structures are the differential cascade voltage switch (DCVS) \cite{Usami} and the current mirror based level shifter (CMLS) \cite{Rajendran,Fassio} as shown in Fig. 1. The DCVS architecture works on the principle of positive feedback to drive the pull up and pull down networks \cite{Usami,Hui_Shao}. This results in a strong contention between the nodes of the circuit. Thus, the contention directly impacts the overall performance 
%%%%%%%%%%%%%%%%%%%%%%%% Fig 1  DCVS & CMLS %%%%%%%%%%%%%%%%%%%%%%%%%%%%%%%%
\begin{figure}[hbtp]
\centering
\includegraphics[scale=0.4]{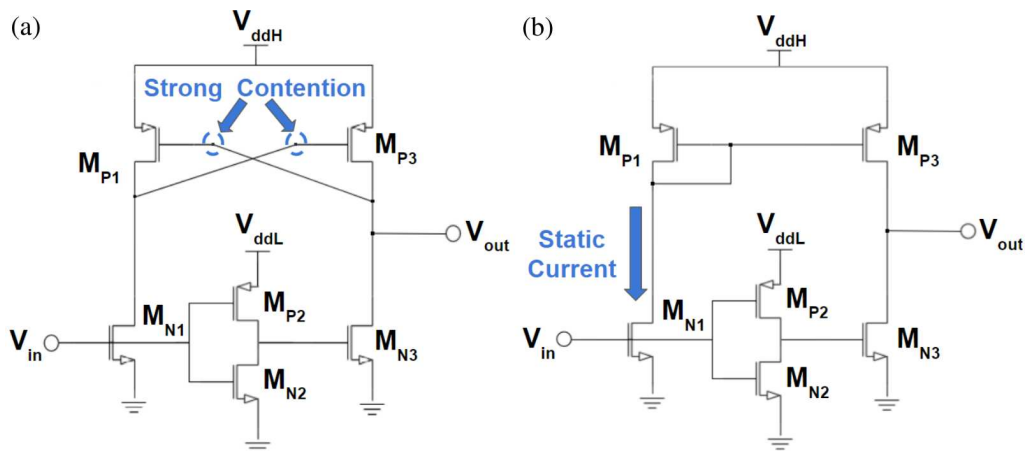}
%\label{Fig. 1(a) Schematic of CMLS.}
\caption{Schematic of conventional (a) differential cascade voltage switch (DCVS) \cite{Usami} and (b) current mirror level shifter (CMLS) \cite{Rajendran}.}
\end{figure}
%%%%%%%%%%%%%%%%%%%%%%%%%%%%%%%%%%%%%%%%%%%%%%%%%%%%%%%%%%%%%%%%%%%%%%%%%%%%%%%%%%%%%%%%%%%%%%%%%%%%%%%%%%%%%%%%%%%%%%%%%%%%%%
 and power dissipation of the LS \cite{Hui_Shao,Hosseini_1,Hasanbegovic}. Moreover, the ultra low voltage operation of DCVS further degrades with lowering supply voltage (V\textsubscript{ddL}) towards the sub-threshold region \cite{Hosseini_1,Hasanbegovic}. Various circuit implementations viz. logic-error correction \cite{Hosseini_1,Osaki,Hui_Shao,kim}, dual threshold voltage \cite{Hasanbegovic,Mohammadi,Matsuzka}, transmission gates \cite{Gak}, clock synchronizers \cite{Chang}, time borrowing techniques \cite{Time_borrowing}, pre-amplifier stage \cite{Matsuzka_2}, cascaded LS \cite{cascaded_LS}, sleep transistors, transistor stacking, and reverse body biasing \cite{Hasanbegovic} are reported in literature which primarily focus on achieving extremely robust LS design for sub-threshold to super-threshold voltage conversion. On the other hand CMLS based LS designs possess a weaker contention and can be used to operate in sub-threshold region while detecting ultra low voltage signals.\\

In this work, we report a modified CMLS based LS design, with the primary goal of achieving ultra low voltage detection while working in the sub-threshold voltage domain. The proposed ultra low power level shifter (ULPLS) architecture is designed using a current limiting PMOS diode, a voltage divider, and an enhanced pull up network. The ULPLS has less Si area overhead since it does not include very long/wide transistors. It can detect a minimum input signal of 70 mV. Besides, the proposed ULPLS design is highly energy efficient, consuming 22.84 nW average power consumption when the input is 0.1 V (at 100 KHz).

%%%%%%%%%%%%%%%%%%%%%%%%%%%%%%%%%%%%%%%%%%%%%%%%%%%%%%%%%%%%%%%%%%%%%%%%%%%%%%%%%%%%%%%%%%%%%%%%%%%%%%%%%%%%%%%%%%%%%%%%%
%%%%%%%%%%%%%%%%%%%%%%%%%%%% Fig. 2 Proposed ULPLS %%%%%%%%%%%%%%%%%%%%%%%%%%%%
\begin{figure}[!ht]
\begin{center}
\includegraphics[scale=0.625]{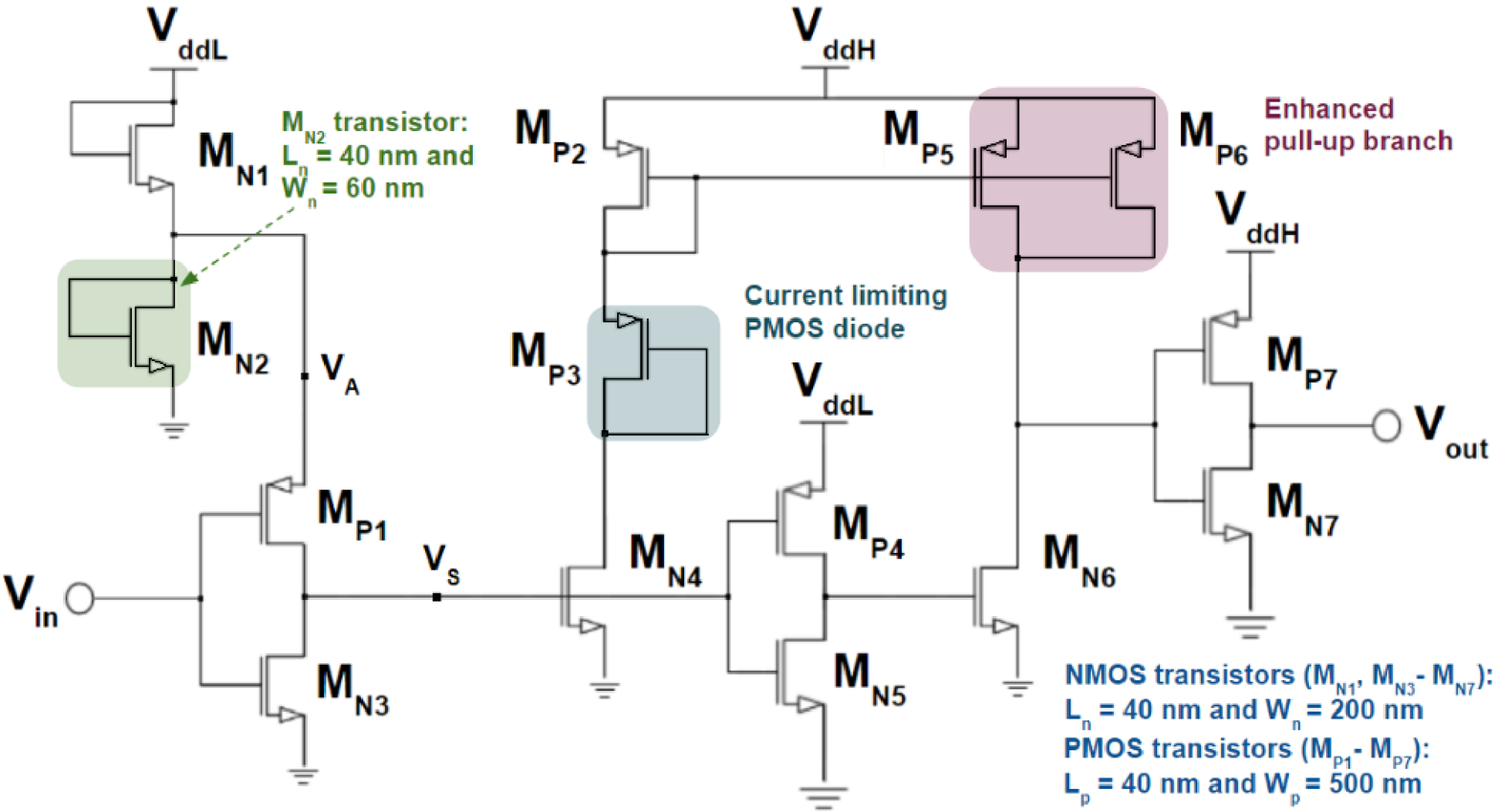}
\label{Fig. 2 Proposed ULPLS}
\caption{Schematic of the proposed ULPLS.}
\end{center}
\end{figure}
%%%%%%%%%%%%%%%%%%%%%%%%%%%%%%%%%%%%%%%%%%%%%%%%%%%%%%%%%%%%%%%%%%%%%%%%%%%%%%%%%%%%%%%%%%%%%%%%%%%%%
\section{METHODOLOGY}
\quad The circuit simulations, in this work, are carried out in SPICE \cite{LTspice} along with the high performance 22 nm Predictive Technology Model (PTM) \cite{PTM}. The default length of transistors are L\textsubscript{NMOS}= L\textsubscript{PMOS}= 40 nm, whereas the default width of NMOS transistor and PMOS transistor are W\textsubscript{NMOS}= 200 nm and W\textsubscript{PMOS}= 500 nm respectively. Moreover, the threshold voltages reported in High Performance 22 nm Predictive Technology Model for NMOS and PMOS transistors are 0.503 V $\&$ 0.460 V \cite{PTM,ASU}. A load capacitance (C\textsubscript{L}) of 200 fF is used which is consistent with literature \cite{Hosseini_1,Osaki}. The average power consumption is calculated considering both the V\textsubscript{ddL} and V\textsubscript{ddH}. Besides, the rise and fall time of the input pulse is taken as 10 ns. As discussed previously, our aim is to design a circuit operating in the sub-threshold region for detection of ultra low voltage signals, without increasing the complexity of fabrication processes. To achieve this, we incorporate changes to the conventional CMLS instead of employing a mixed-threshold voltage scheme \cite{Hosseini_1,Gak,Fassio}.

\section{RESULTS AND DISCUSSION}
\subsection{Proposed Design}
\quad Voltage levels scaled down to sub-threshold region i.e. comparable to V\textsubscript{th} may lead to various reliability issues in DCVS \cite{Hasanbegovic,Fassio}. This is avoided in CMLS design, where a weak contention exists due to the weakly-ON PMOS transistor (M\textsubscript{P1}) \cite{kim,Fassio}. However, there is a large static power dissipation through the M\textsubscript{P1} and M\textsubscript{N1} transistors \cite{Hosseini_1,Hasanbegovic,Late}. To largely reduce the static current in this path, several design modification have been reported in literature utilizing cascode current mirror \cite{Fassio,Prashant}, Wilson current mirror \cite{Lutkemier,Hosseini_1}, back biasing \cite{Prashant}, body biasing \cite{Late} and current limiting elements \cite{Fassio,Hosseini_1,Hasanbegovic,kim}.\\

In \cite{Hosseini_1}, Hosseini et al. reported one such high performance CMLS. In this design a modified Wilson current mirror is used along with a diode-connected PMOS transistor (M\textsubscript{P4}) and an additional pull down network (M\textsubscript{N3}) to further weaken the contention \cite{Hosseini_1}. Nonetheless, the drawback of such designs is the usage of large size transistors specially M\textsubscript{P1} \cite{Mohammadi,Prashant,Late}. To address these concerns, we propose an ultra low power level shifter by modifying the conventional CMLS \cite{Fassio} as shown in Fig. 2. We employ a current limiting PMOS diode (M\textsubscript{P3}) with an enhanced pull up network (M\textsubscript{P5} \& M\textsubscript{P6}) to reduce the power dissipation and improve its transition time. We have also included a voltage divider system (M\textsubscript{N1} \& M\textsubscript{N2}) that helps weakening the contention by not allowing the pull up network to partially turn on. Apart from that we have implemented a dual supply voltage design with V\textsubscript{ddH} and V\textsubscript{ddL} as 0.8 V and 0.4 V respectively.

\subsection{Low voltage detection of proposed ULPLS}
\quad The aim of our work is to design a LS for ultra low voltage emerging applications. Thus, we need to detect input voltage signals (V\textsubscript{in}) around the sub-threshold range, in order to attain high energy efficiency. In Table \ref{Vout} and Table \ref{Vinl}, we demonstrate the effect of V\textsubscript{in} and V\textsubscript{ddL} variations on the performance of the ULPLS.
%%%%%%%%%%%%%%%%%%%%%%%% Table 1 Vddl Selection  %%%%%%%%%%%%%%%%%%%%%%%%%%%%%
\begin{table}[!ht]
\caption{Variation of V\textsubscript{out} of the ULPLS with variation in V\textsubscript{ddL}}
\begin{center}
\label{Vout}
\begin{tabular}[h]{p{0.45cm}p{0.45cm}p{1cm}p{1.75cm}p{1.85cm}p{0.85cm}p{1.75cm}}
\hline
\footnotesize{V\textsubscript{in} \newline (V)}
&\footnotesize{V\textsubscript{ddL} (V)}
&\footnotesize{V\textsubscript{out,High} (V)}
&\footnotesize{V\textsubscript{out,Low} (nV)}
&\footnotesize{Avg. \newline Power (nW)} &\footnotesize{T\textsubscript{D Max.} (ns)}  &\footnotesize{PDP ($\times10^{-18}$J)}\\
\hline\hline
\footnotesize{0.5} &\footnotesize{0.5} &\footnotesize{0.8} 
&\footnotesize{$\sim458\times10^{3}$} &\quad\footnotesize{71.97} &\footnotesize{436}
&\footnotesize{$\sim31375.4$}\\

\footnotesize{0.4} &\footnotesize{0.4} &\footnotesize{0.8} 
&\footnotesize{$\sim253.5$} &\quad\footnotesize{19.88} &\footnotesize{460}
&\footnotesize{$\sim9151.35$}\\

\footnotesize{0.3} &\footnotesize{0.3} &\footnotesize{0.8} 
&\footnotesize{$\sim255$} &\quad\footnotesize{19.22} &\footnotesize{982}
&\footnotesize{$\sim18869.3$}\\

\footnotesize{0.2} &\footnotesize{0.2} &\footnotesize{0.8} 
&\footnotesize{$\sim270$} &\quad\footnotesize{32.95} &\footnotesize{2180}
&\footnotesize{$\sim71853.7$}\\
\hline
\end{tabular}
\end{center}
\end{table}
%%%%%%%%%%%%%%%%%%%%%%%% Table 2 Vinl Selection  %%%%%%%%%%%%%%%%%%%%%%%%%%%%%
\begin{table}[H]
\caption{Variation of V\textsubscript{out} of the ULPLS with variation in V\textsubscript{in}}
\label{Vinl}
\begin{center}
\begin{tabular}[h]{p{1.5cm}p{0.45cm}p{1cm}p{1.75cm}p{1.85cm}p{0.85cm}p{1.75cm}}
\hline
\footnotesize{V\textsubscript{in} \newline (V)}
&\footnotesize{V\textsubscript{ddL} (V)}
&\footnotesize{V\textsubscript{out,High} (V)}
&\footnotesize{V\textsubscript{out,Low} (nV)}
&\footnotesize{Avg. \newline Power (nW)} &\footnotesize{T\textsubscript{D Max.} (ns)}  &\footnotesize{PDP ($\times10^{-18}$J)}\\

\hline\hline

\footnotesize{0.4} &\footnotesize{0.4} &\footnotesize{0.8} 
&\footnotesize{$\sim254$} &\quad\footnotesize{19.88} &\footnotesize{460}
&\footnotesize{$\sim9151.35$}\\

\footnotesize{0.2} &\footnotesize{0.4} &\footnotesize{0.8} 
&\footnotesize{$\sim254$} &\quad\footnotesize{22.49} &\footnotesize{457}
&\footnotesize{$\sim10276.8$}\\

\footnotesize{0.1} &\footnotesize{0.4} &\footnotesize{0.8} 
&\footnotesize{$\sim254$} &\quad\footnotesize{22.62} &\footnotesize{423}
&\footnotesize{$\sim9561.17$}\\

\footnotesize{$80\times10^{-3}$} &\footnotesize{0.4} &\footnotesize{0.8} 
&\footnotesize{$\sim254$} &\quad\footnotesize{29.73} &\footnotesize{382}
&\footnotesize{$\sim11348.4$}\\

\footnotesize{$70\times10^{-3}$} &\footnotesize{0.4} &\footnotesize{0.8} 
&\footnotesize{$\sim254$} &\quad\footnotesize{31.26} &\footnotesize{526}
&\footnotesize{$\sim16428.9$}\\
\hline
\end{tabular}
\end{center}
\end{table}
The output waveforms obtained are shown in Fig. 3. A detailed comparison of our proposed ULPLS is made with state of the art CMLS architectures reported in literature (as shown in Table \ref{Comparison Table}).
%%%%%%%%%%%%%%%%%%%%%%%% Fig 3 IP OP waveform %%%%%%%%%%%%%%%%%%%%%%%%%%%%%%%%
\begin{figure}[H]
\centering
\includegraphics[scale=0.535]{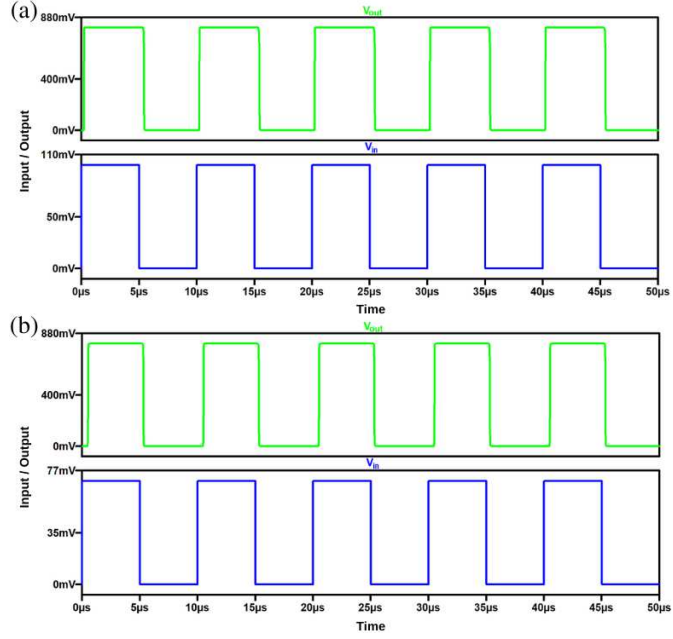}
%\label{Fig. 1(a) Schematic of CMLS.}
\caption{Transient response of proposed ULPLS operating at 100 KHz with (a) V\textsubscript{in} = 0.1 V \& V\textsubscript{ddL} = 0.4 V and (b) V\textsubscript{in} = 70 mV \& V\textsubscript{ddL} = 0.4 V. The high supply voltage is maintained at 0.8 V.}
\end{figure}
%%%%%%%%%%%%%%%%%%%%%%%%%%%%%%%%%%%%%%%%%%%%%%%%%%%%%%%%%%%%%%%%%%%%%%%%%%%%%%%%%%%%%%%%%%%%%%%%%%%%%%%%%%%%%%%%%%%%%%%%%%%%%% 

%%%%%%%%%%%%%%%%%%%%%%%%%%%% Table 3 Benchmark Table %%%%%%%%%%%%%%%%%%%%%%%%%%%%%%%%%%%%%%%%%%%%%%%%%%%%
\begin{table}[H]
\begin{center}
\caption{Comparison with state of the art LS architectures}
\begin{tabular}[h]{p{0.75cm}p{0.65cm}p{0.65cm}p{0.65cm}p{2.5cm}p{3.15cm}p{0.375cm}}
\hline
\footnotesize{Level shifter} 
&\footnotesize{Tech. \newline (nm)} 
&\footnotesize{V\textsubscript{ddL} (V)}
&\footnotesize{V\textsubscript{ddH} (V)}
&\footnotesize{Min. V\textsubscript{in} (mV), at f\textsubscript{oper.}(Hz)}
&\footnotesize{Avg. Power (nW), at f\textsubscript{oper.}(Hz)}
&\footnotesize{T\textsubscript{Total}\textsuperscript{$\dagger$}}\\
\hline\hline
%%%%%%%%%%%%%% High Frequency %%%%%%%%%%%%
%  \footnotesize{\cite{Rajendran}} 
% &\footnotesize{600} &\footnotesize{1.8} &\footnotesize{1.8} &\footnotesize{580,$100\times$10\textsuperscript{6}}
% &\footnotesize{4200,$100\times$10\textsuperscript{6}} 
% &\footnotesize{7}\\
%%%%%%%%%%%%%%%%%%%%%%%%%%%%%%%%%%%%%%%%%%%%%%
\footnotesize{\cite{Fassio}} 
&\footnotesize{180} &\footnotesize{0.4} &\footnotesize{1.8}
&\footnotesize{85, 100}
 &\footnotesize{6.9, $100\times$10\textsuperscript{3}}
 &\footnotesize{11}\\

 \footnotesize{\cite{Hosseini_1}} 
 &\footnotesize{180} &\footnotesize{0.4} &\footnotesize{1.8}
 &\footnotesize{320,10\textsuperscript{6}}
&\footnotesize{680, 10\textsuperscript{6}}
&\footnotesize{9}\\

%  \footnotesize{\cite{Hosseini_2}} 
%  &\footnotesize{180} &\footnotesize{0.45} &\footnotesize{1.8}
%  &\footnotesize{360,10\textsuperscript{6}}
%  &\footnotesize{175, 10\textsuperscript{6}}
% &\footnotesize{16}\\
  
\footnotesize{\cite{Kabirpour}} 
&\footnotesize{180} &\footnotesize{0.4} &\footnotesize{1.2}
&\footnotesize{50, $10\times$10\textsuperscript{3}}
&\footnotesize{76.34, 10\textsuperscript{6}}
 &\footnotesize{11}\\

\footnotesize{\cite{Magh}} 
&\footnotesize{180} &\footnotesize{0.4} &\footnotesize{1.8}
&\footnotesize{330, $200\times$10\textsuperscript{3}}
&\footnotesize{61.5, $500\times$10\textsuperscript{3}}
 &\footnotesize{13}\\

%\hline
\footnotesize{This-work} 
&\footnotesize{22} &\footnotesize{0.4} &\footnotesize{0.8}
&\footnotesize{70, $100\times$10\textsuperscript{3}}
&\footnotesize{22.84, $100\times$10\textsuperscript{3}}
 &\footnotesize{14}\\
\hline
\multicolumn{5}{c}{\footnotesize{ $\dagger$ T\textsubscript{Total} represents the transistor count.}} 
\label{Comparison Table}
\end{tabular}
\end{center}
\end{table}
The LS architectures reported in \cite{Fassio} and \cite{Kabirpour} detect 85 mV input at f\textsubscript{oper.} (operating frequency) = 100 Hz and 50 mV input at f\textsubscript{oper.} = 10 KHz respectively. However, it is important to detect ultra low voltage at higher frequencies while maintaining high energy efficiency. In \cite{Hosseini_1}, the LS can  detect voltages in the range of 320 mV at f\textsubscript{oper.} = 1000 KHz with an avg. power of 680 nW. Apart from that, in \cite{Magh} some improvement in avg. power is shown with a similar input detection of 330 mV at f\textsubscript{oper.} = 200 KHz. Thus, we find the proposed ULPLS emerges as a suitable candidate to detect ultra low voltage at significantly large frequency. It detects 70 mV input signal at f\textsubscript{oper.} = 100 KHz with an avg. power of 22.84 nW. Although the avg. power consumption is slightly larger for the proposed ULPLS compared to the LS reported in \cite{Fassio}, we find it is notably less than the power consumed by other architectures (e.g. level shifters of \cite{Hosseini_1}, \cite{Kabirpour}, and \cite{Magh}) available in literature.
%%%%%%%%%%%%%%%%%%%%%%%%%%%%%%%%%%%%%%%%%%%%%%%%%%%%%%%%%%%%%%%%%%%%%%%%%%%
\subsection{Impact of Temperature Variation on Performance}
\quad Temperature is one of the important environmental parameters that largely impact the circuit performance \cite{Ajit}. Since the temperature variation can lead to non-linear distortion, determining the robustness of the ULPLS at any specific operating temperature becomes crucial \cite{Fassio,Kabirpour}. In Table \ref{tab:temps}, we show the impact of temperature variation on the operation of the proposed ULPLS. The circuit shows robust performance, with complete voltage swing at output across the temperature range of -40 $^{\circ}$C to 125 $^{\circ}$C. Besides, the change in average power consumption is small (21.62 nW to 32.17 nW) when the temperature is largely varied within the range -40 $^{\circ}$C to 125 $^{\circ}$C.  
%%%%%%%%%%%%%%%%%%%%%%%%%%%%%%%%%%% Table 4 Temp Sweep %%%%%%%%%%%%%%%%%%%%%%%%%%%%%%
\begin{table}[!h]
        \centering
       \caption{Temperature sweep analysis of ULPLS}
        \begin{tabular}[h]{p{0.75cm}p{2.2cm}p{2cm}p{1cm}p{1.5cm}}
        \hline
      \footnotesize{Temp. \newline ($^{\circ}$C)} &\footnotesize{Voltage Swing (mV)} &\footnotesize{Avg. Power (nW)} &\footnotesize{T\textsubscript{D,Max.} (ns)} &\footnotesize{PDP ($\times10^{-18}$J)}\\
        \hline \hline
        
        -40 &\footnotesize{$\sim$800}& \footnotesize{25.30} & \footnotesize{536} &\footnotesize{$\sim13561$}  \\
        
        0 &\footnotesize{$\sim$800} & \footnotesize{21.62} & \footnotesize{647} & \footnotesize{$\sim13988$} \\
        
        27 & \footnotesize{$\sim$800} & \footnotesize{22.62} & \footnotesize{423} & \footnotesize{$\sim9568$}  \\
        
        125 & \footnotesize{$\sim$800} & \footnotesize{32.17} & \footnotesize{106} & \footnotesize{$\sim3410$}  \\
        \hline
        \end{tabular}
       \label{table 3 Temp Sweep}
     \label{tab:temps}
\end{table}
%%%%%%%%%%%%%%%%%%%%%%%%%%%%%%%%%%%%%%%%%%%%%%%%%%%%%%%%%%%%%%%%%%%%%%%%%%%
\subsection{Impact of Process Variation on Performance}
\begin{figure}[!h]
\centering
\includegraphics[scale=0.435]{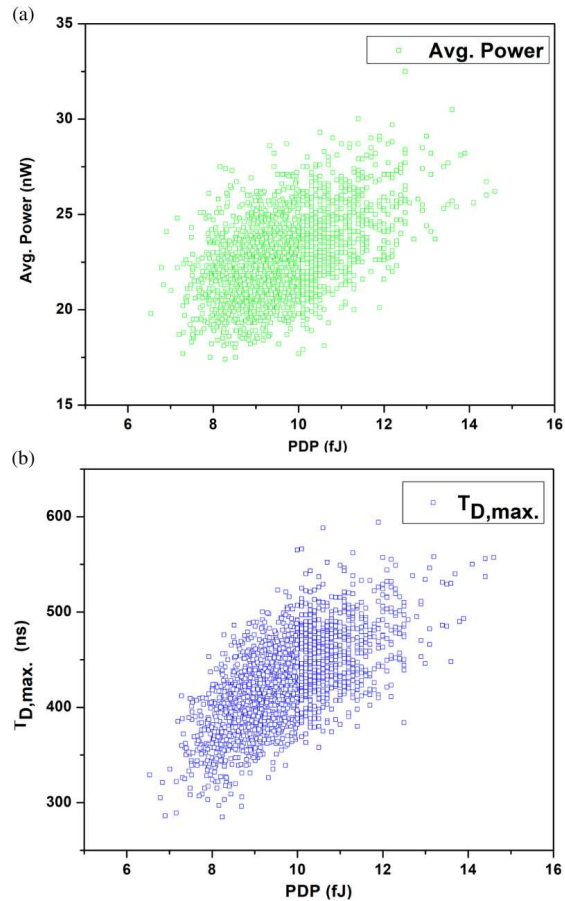}
\caption{ULPLS performance with process variation, where (a) power and (b) delay plots are shown with 10$\%$ tolerance on the supply voltages (V\textsubscript{ddL} \& V\textsubscript{ddH}) and 4$\%$ tolerance on transistor sizing.}
\label{Fig_MC}
\end{figure}

\begin{figure}[!h]
\centering
\includegraphics[scale=0.535]{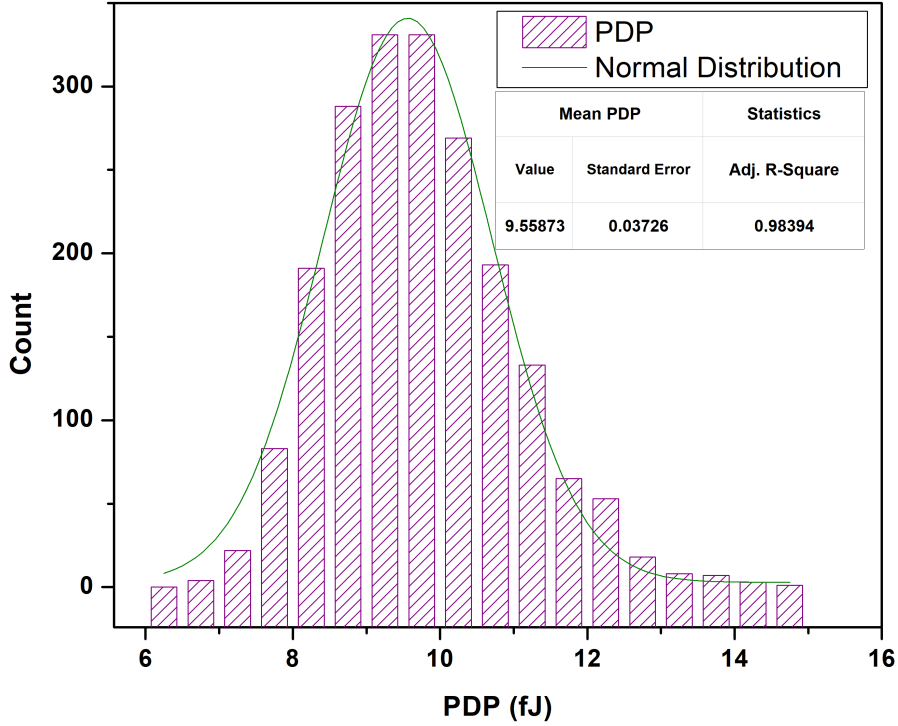}
\caption{Histogram of Proposed ULPLS PDP with process variation of 10$\%$ tolerance on the supply voltages (V\textsubscript{ddL} \& V\textsubscript{ddH}) and 4$\%$ tolerance on transistor sizing.}
\label{Fig:MC_PDP}
\end{figure}

\quad Non-uniform process conditions can led to a variation in the transistors characteristics \cite{Ajit}. The changes in the features like transistor sizing and power supply cause variation in the performance of any electrical circuit. To probe into these variations which are usually uncorrelated, we take the help of Monte Carlo analysis. The four PVT corners shown in Fig. \ref{Fig_MC} are a result of a 10 $\%$ tolerance on the supply voltages ($V_{ddH}$ \& $V_{ddL}$) and 4 $\%$ tolerance on the transistor sizing. The variation in Power Delay Product (PDP) of the proposed LS with the prior mentioned process variation is shown in Fig. \ref{Fig:MC_PDP}. 
%%%%%%%%%%%%%%%%%%%%%%%% Fig 5 MC Analysis %%%%%%%%%%%%%%%%%%%%%%%%%%%%%%%%
%%%%%%%%%%%%%%%%%%%%%%%%%%%%%%%%%%%%%%%%%%%%%%%%%%%%%%%%%%%%%%%%%%%%%%%%%%%

%%%%%%%%%%%%%%%%%%%%%%%% Fig 5 MC Analysis %%%%%%%%%%%%%%%%%%%%%%%%%%%%%%%%
%%%%%%%%%%%%%%%%%%%%%%%%%%%%%%%%%%%%%%%%%%%%%%%%%%%%%%%%%%%%%%%%%%%%%%%%%%%
%%%%%%%%%%%%%%%%%%%%%%%%%%%%%%%%%%%%%%%%%%%%%%%%%%%%%%%%%%
\subsection{Worst Case Analysis: Sizing M\textsubscript{N1} \& M\textsubscript{N2}}
\begin{figure}[H]
\centering
\includegraphics[scale=0.625]{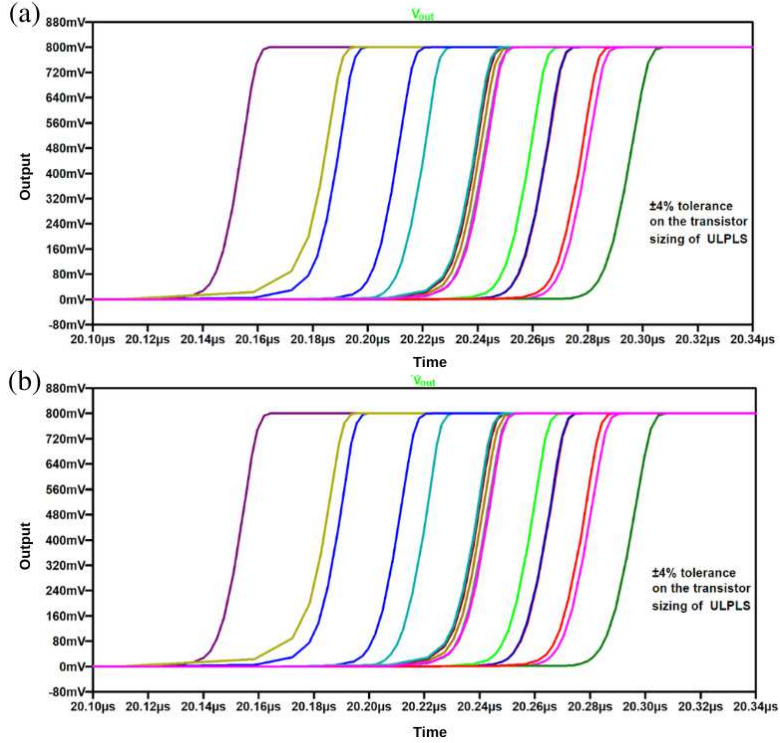}
%\label{Fig. 1(a) Schematic of CMLS.}
\caption{(a) Rise and (b) fall time of output waveforms via worst case analysis with 4 $\%$ tolerance on M\textsubscript{N1} and M\textsubscript{N2} with V\textsubscript{ddH} = 0.8 V \& V\textsubscript{ddL} = 0.4 V.}
\label{Fig_WC}
\end{figure}

\quad As discussed earlier, M\textsubscript{N1} \& M\textsubscript{N2} play a key role in determining the lowest possible detectable signal ($V_{in}$). Thus, the reliability of the design is strongly influenced by sizing ratio of these transistors. By performing a worst case analysis on the transistor sizing, we can account for the reliability of the circuit under the worst possible conditions \cite{WC_1,WC_2}. The following analysis is done with a tolerance factor of 4 $\%$ on the sizing of the transistors as shown in Fig. \ref{Fig_WC}.
%%%%%%%%%%%%%%%%%%%%%%%% Fig 4 WC Analysis %%%%%%%%%%%%%%%%%%%%%%%%%%%%%%%%
%%%%%%%%%%%%%%%%%%%%%%%%%%%%%%%%%%%%%%%%%%%%%%%%%%%%%%%%%%%%%%%%%%%%%%%%%%%%%%%%%%%%%%%%%%%%%%%%%%%%%%%%%%%%%%%%%%%%%%%%%%%%%%
\section{Conclusion}
\quad In this paper, an ultra low-power level shifter is proposed that works effectively in sub-threshold region. By incorporating a current limiting PMOS diode, enhanced pull up network, and a voltage divider to the standard CMLS architecture, a minimum detectable input voltage of 70 mV has been observed. The proposed ULPLS detects a significantly low voltage ($\sim$ 70 mV) with high energy efficiency. The average power consumption of the ULPLS is $\sim$ 22.84 nW while operating at 100 KHz. Besides, this LS design exhibits a tolerance of 4 $\%$ on the sizing of the transistors. Moreover considering supply voltage and temperature variation, we find that the proposed ULPLS shows robust circuit performance. This study may further be extended to determine the reliability of the proposed architecture and tolerance to noise.

% \section*{Acknowledgment}
% \quad The authors thank the Department of Electronics and Telecommunication Engineering, Indian Institute of Engineering Science and Technology, Shibpur for the computing facilities and resources.

\end{document}